\documentclass[12pt]{article}
\usepackage{graphicx}
\usepackage{amsmath}
\usepackage{subfigure}

\begin{document}
\begin{titlepage}
\begin{description}
\item[Title:] 
Noise-Induced Sampling of Alternative Hamiltonian Paths in
Quantum Adiabatic Search\\
\item[Author:] Frank Gaitan\\
               Department of Physics\\
               Southern Illinois University\\
               Carbondale, IL 62901-4401\\
\item[Pages:] $11$ pages of text (double-space)
\item[Figures:] $5$ figures
\item[Tables:] $4$ tables
\end{description}

\begin{center}
\textbf{Technical Abstract}
\end{center}
We numerically simulate the effects of noise-induced sampling of alternative
Hamiltonian paths on the ability of quantum adiabatic search (QuAdS) to solve
randomly-generated instances of the NP-Complete problem N-bit Exact Cover 3. 
The noise-averaged median runtime is determined as noise-power and number 
of bits $N$ are varied, and power-law and exponential fits are made to the 
data. Noise is seen to slowdown QuAdS, though a downward shift in the scaling 
exponent is found for $N>12$ over a range of noise-power values. We discuss 
whether this shift might be connected to arguments in the literature that 
suggest that altering the Hamiltonian path might benefit QuAdS performance.\\

\begin{center}
\textbf{Non-Technical Abstract}
\end{center}
We numerically simulate the effects of noise on the ability of the Quantum
Adiabatic Search (QuAdS) algorithm to solve randomly-generated instances of
the NP-Complete problem N-bit Exact Cover 3. The noise-averaged median runtime
is determined as noise-power and number of bits N are varied, and power-law
and exponential fits are made to the data. Noise is seen to slowdown QuAdS,
though a downward shift in the scaling exponent is found for $N > 12$ over a
range of noise-power values. We discuss whether this shift might be connected 
to arguments in the literature that suggest that altering the path followed by 
the search Hamiltonian might benefit QuAdS performance.

\begin{description}
\item[Keywords:] quantum adiabatic search; quantum algorithms; computational
complexity; NP-Completeness; noise
\end{description}
\end{titlepage}
\pagebreak

\section{Introduction}
\label{sec1}

One of the deepest open questions in theoretical computer science is whether
the computational complexity classes $P$ and $NP$ are equal \cite{gar+john}.
It is widely conjectured that these two classes are different, though it is
known that should a polynomial-time algorithm be found for an $NP$-Complete 
problem, then $P=NP$. In 2001, Farhi et.\ al.\ \cite{far1} examined whether a 
quantum algorithm might be able to solve an $NP$-Complete problem in 
polynomial-time. They used the quantum adiabatic search (QuAdS) algorithm
\cite{far2} to find solutions to randomly-generated hard instances of the
NP-Complete problem $N$-bit Exact Cover 3 which they believed to be classically
intractable for sufficiently large $N$. Using a digital computer they 
numerically simulated the QuAdS dynamics. They determined the algorithm's
median runtime to solve this class of instances for a restricted range of $N$ 
values and found their results could be fit with a quadratic scaling relation. 
It was noted that should classical algorithms truly require exponential time 
to solve this class of instances, and should the quadratic scaling behavior of 
QuAdS persist to large $N$, then QuAdS could outperform classical algorithms 
on this class of instances, though not necessarily on the worst case instances.

QuAdS works well so long as the quantum dynamics is adiabatic. This requires
the runtime $T$ to be large compared to $1/\Delta^{2}$, where $\Delta$ is the
smallest value (encountered during the dynamical evolution) of the energy gap
between the ground and first-excited states. When $\Delta$ is too small, the
adiabatic condition is violated, and QuAdS performance suffers. Farhi and
co-workers \cite{far3} examined a case where a failure of QuAdS was transformed
into a success if the path followed by the search Hamiltonian $H(t)$ differed
from the Hamiltonian path used in Refs.~\cite{far1} and \cite{far2} that 
linearly interpolates from an initial to a final Hamiltonian. The essential 
point is that, should the linearly interpolating search Hamiltonian
produce a $\Delta$ that is too small, varying the Hamiltonian path may cause 
the new search Hamiltonian to produce a larger $\Delta$ and thus improve QuAdS
performance.

A number of papers have considered the robustness of QuAdS performance to
noise \cite{childs}--\cite{fg}. In this paper we extend the simulations 
reported in Ref.~\cite{fg} in two important ways. First, the simulation
results presented here examine QuAdS performance in the presence of non-uniform
noise in which each qubit interacts with a different noise field. 
Ref.~\cite{fg} focused on uniform noise. Second, the simulations in this 
paper are done at larger noise power and larger numbers of qubits $N$. These
differences allow the simulations in this paper to sample a larger range of 
Hamiltonian path variations than was possible in Ref.~\cite{fg}, and so 
provide a better opportunity to explore how Hamiltonian path variation 
impacts QuAdS performance.

This paper is organized as follows: (i)~Section~\ref{sec2} briefly summarizes 
the QuAdS algorithm, our noise model, and the simulation protocol;
(ii)~Section~\ref{sec3} presents our simulation results; and 
(iii)~Section~\ref{sec4} closes with a discussion of these results.

\section{Background}
\label{sec2}

We begin with a description of the $NP$-Complete problem $N$-bit Exact 
Cover~3 (EC3) \cite{gar+john,far1}. An instance of EC3 is specified by a set 
of clauses $C_{i}$, with $i=1,\ldots , L$, and each clause $C_{i}$ is specified 
by $3$ integers: $C_{i}=(a(i),b(i),c(i))$. The integers $a(i)$, $b(i)$, and 
$c(i)$ satisfy $a(i)<b(i)<c(i)$ and take values in the range [$1,\ldots , N$].
Generally, the number of clauses $L$ varies from one EC3 instance to another. 
A binary vector $z=(z_{1},\ldots , z_{N})$ (with $z_{j}=0,1$) satisfies the 
clause $C_{i}$ if its components $z_{a(i)}$, $z_{b(i)}$, and $z_{c(i)}$ 
satisfy $z_{a(i)}+z_{b(i)}+z_{c(i)}=1$. Otherwise, $z$ is said to violate 
$C_{i}$. A binary vector $z$ solves an instance of EC3 if it satisfies all of 
its clauses. Finally, an EC3 instance is said to have a unique satisfying 
assignment (USA instance) if only one binary vector solves it. 

In QuAdS an $n$-qubit register is initially prepared in the groundstate of a
Hamiltonian $H_{i}$. The only conditions placed on $H_{i}$ are that its 
groundstate be non-degenerate and easy to prepare. $H_{i}$ is then 
adiabatically evolved over a time $T$ into a final Hamiltonian $H_{P}$. The 
final Hamiltonian is constructed so that a basis for its groundstate 
eigenspace encodes all solutions to the instance of the computational problem 
that is to be solved. The details of how $H_{P}$ is  constructed from an 
instance of EC3 are described in Ref.~\cite{far1}. The construction of $H_{i}$ 
and $H_{P}$ are such that both are dimensionless and have energy-level spacing 
$\Delta E\sim 1$. Since the initial state is the groundstate of $H_{i}$, the 
adiabatic dynamics insures that the final state will be in the groundstate
eigenspace of $H_{P}$ with probability $P\rightarrow 1$ as $T\rightarrow
\infty$. An appropriate measurement of the quantum register at the end of
the adiabatic evolution then yields one of the instance solutions. In
Refs.~\cite{far1} and \cite{far2} the time-dependent Hamiltonian $H(t)$ that
drives the QuAdS dynamics linearly interpolates from $H_{i}$ to $H_{P}$,
\begin{equation}
\label{Hnon}
H(t)=\left( 1-\frac{t}{T}\right)H_{i} +\left(\frac{t}{T}\right)H_{P}
 \hspace{0.1in} ,
\end{equation}
where $0\leq t\leq T$, and $T$ is sufficiently large that $H(t)$ produces
adiabatic dynamics. The simulations in Ref.~\cite{far1} randomly generated $75$
USA instances of $N$-bit EC3 which were believed to be hard instances for both
classical algorithms and QuAdS. The median runtime $\overline{T}(N)$ for QuAdS
to succeed on these instances was found for $7\leq N\leq 20$. It was found that
the simulation results could be fit with a quadratic scaling relation 
$\overline{T}(N)\sim N^{2}$.

To study the impact of noise on QuAdS we introduce classical noise fields
$\mathbf{N}_{j}(t)$ ($j=1,\ldots ,N$) that couple to the qubits via the
Zeeman interaction
\begin{equation}
\label{Hint}
H_{int}(t)=-\sum_{j=1}^{N}\,\mbox{\boldmath $\sigma$}_{j}\cdot\mathbf{N}_{j}
                (t) \hspace{0.1in} .
\end{equation}
In this paper we focus on non-uniform noise where each qubit is acted on by a
different noise field: $\mathbf{N}_{j}(t)\neq \mathbf{N}_{i}(t)$ ($j\neq i$).
A detailed presentation of our noise model is given in Ref.~\cite{fg}---we
summarize its essential properties here. Each noise field $\mathbf{N}_{j}(t)$ 
is a sequence of randomly occurring fluctuations with profile 
$\mathbf{F}_{j}(t)$:
\begin{equation}
\mathbf{N}_{j}(t) = \sum_{k=1}^{N_{f}}\,\mathbf{F}_{j}(t-t_{k})
      \hspace{0.5in} (j=1,\ldots , N) \hspace{0.1in} .
\end{equation}
Here $t_{k}$ is the temporal center of the $k$-th fluctuation and $N_{f}$ is
the number of fluctuations. The fluctuations have the following statistical
properties: (1)~the number of fluctuations $N_{f}$ that occur in a time $T$ is 
Poisson distributed with average fluctuation rate $\overline{n}$; (2)~each 
fluctuation profile $\mathbf{F}_{j}(t-t_{k})$ is a square pulse with height 
$x_{j,k}$ and temporal width $2\tau$, where $\tau$ is the thermal relaxation 
time; (3)~the height $x_{j,k}$ is Gaussian distributed with zero mean and 
variance $\sigma^{2}$; and (4)~the times $t_{k}$ are uniformly distributed 
over [$0,T$]. The simulations allow the polarization of $\mathbf{F}_{j}(t)$ to
be either: (i)~fixed along $\hat{\mathbf{x}}$, $\hat{\mathbf{y}}$, or
$\hat{\mathbf{z}}$; or (ii)~to fluctuate simultaneously along all $3$ 
directions. In Ref.~\cite{fg} it was found that noise polarized along
$\hat{\mathbf{y}}$ caused the largest slowdown of QuAdS and so we focus on 
$y$-polarized noise throughout this paper. The time-averaged noise power
$\overline{P}$ is related \cite{fg} to $\overline{n}$, $\sigma^{2}$,
and $\tau$ via $\overline{P}=2\overline{n}\sigma^{2}\tau$. The simulations
described below use $\sigma =0.2$; $\tau = 1$; and average noise power in the
range $0.001\leq \overline{P}\leq 0.013$. The average fluctuation rate is then
determined from $\overline{n}=\overline{P}/2\sigma^{2}\tau$.

The QuAdS simulation protocol with noise is described in Ref.~\cite{fg}. As
with noiseless QuAdS, the protocol with noise begins by producing $75$
randomly generated USA instances of $N$-bit EC3. The simulations described
below were done for $7\leq N\leq 16$. For each USA instance, $10$ noise
environments $\{\mathbf{N}_{j}^{m}(t): j=1,\ldots ,N; \; m= 1,\ldots ,10\}$
were generated. For each $m$, $H_{int}(t)$ is determined by plugging the 
$\{\mathbf{N_{j}}^{m}(t)\}$ into eq.~(\ref{Hint}). The noiseless QuAdS
Hamiltonian $H(t)$ is given by eq.~(\ref{Hnon}), where $H_{i}$ is the same for 
all USA instances and each USA instance determines its own $H_{P}$ 
\cite{far1}. The noisy QuAdS Hamiltonian $\mathcal{H}(t)$ is then
\begin{equation}
\label{Htot}
\mathcal{H}(t) = H(t)+H_{int}(t) \hspace{0.1in} .
\end{equation}
For each USA instance and noise environment, $\mathcal{H}(t)$ drives the
Schrodinger dynamics of QuAdS. This dynamics is numerically simulated to
find the runtime for QuAdS to succeed on that instance and that noise 
environment. For each $N$, a total of $750=75\times 10$ QuAdS runtimes are
generated. The noise-averaged median runtime $\langle\overline{T}(N)\rangle$
is then identified with the median of the $750$ runtimes. We determined the
best power-law and exponential fits to the simulation results, and
calculate their associated $\chi^{2}_{fit}$ and probability $P(\chi^{2}>
\chi^{2}_{fit})$. The simulations were done on the National Science
Foundation TeraGrid cluster. 

\section{Results}
\label{sec3}

\begin{figure}[p!]
\centerline{\includegraphics[scale=0.7]{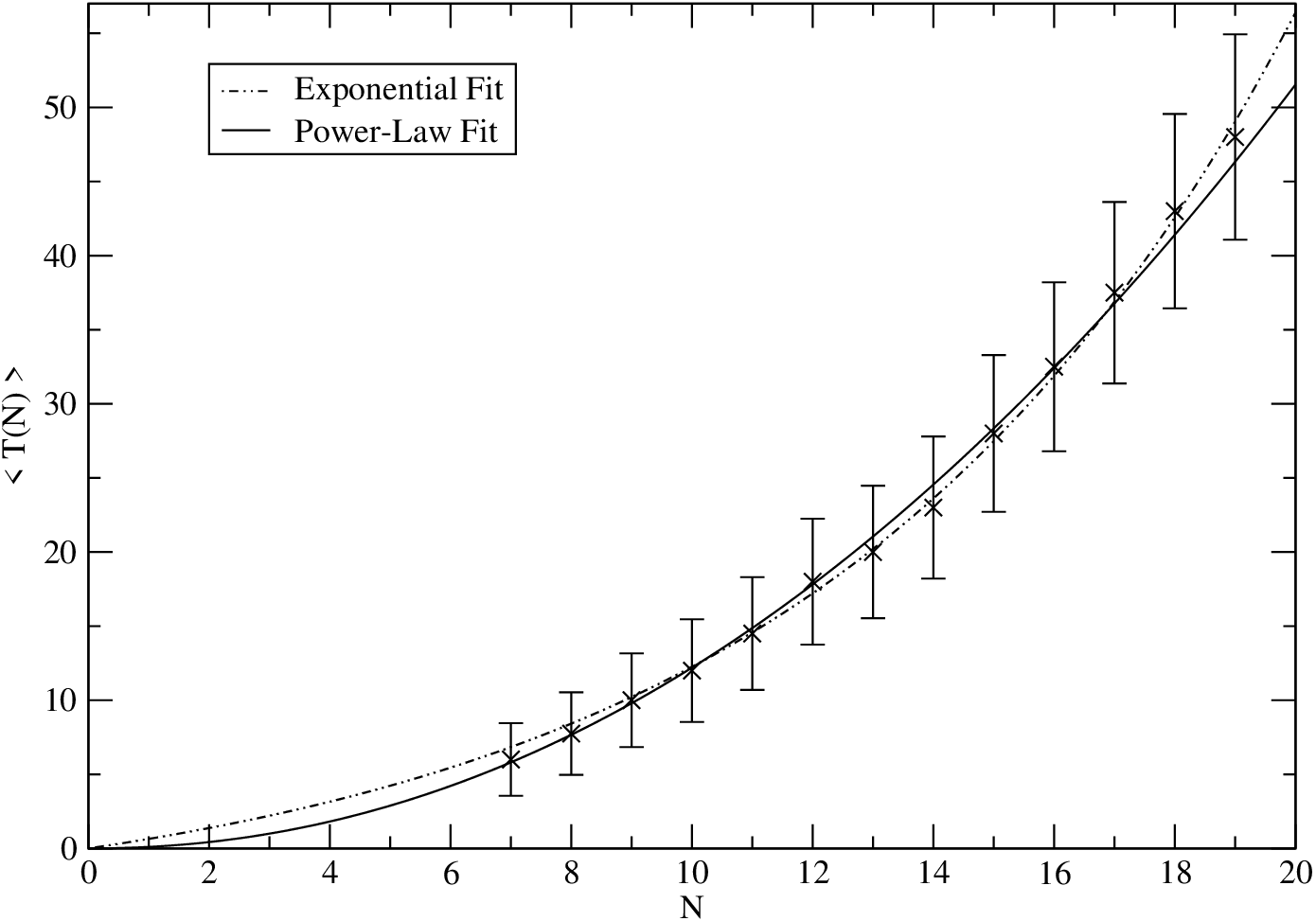}}
\vspace{3.5in}
\caption{\label{figure1}}
\end{figure}

We now present our simulation results. As a baseline for the noisy simulations
we repeat the noiseless calculation of Ref.~\cite{far1} for $7\leq N\leq 19$. 
Our results appear in Figure~\ref{figure1} which contains best power-law and
exponential fits to the data. The power-law fit $\overline{T}(N)=aN^{b}$
has fit parameters $a=0.1016$ and $b=2.079$. The value of $\chi^{2}$ for the
fit is $\chi^{2}_{fit}=0.321$ and the probability $P(\chi^{2}>\chi^{2}_{fit})
=0.9999$. The closer this probability is to $1$, the more consistent the
data-set is with the fitting function. The exponential fit $\overline{T}(N)=a
[\exp (bN)-1]$ has: (i)~fit parameters $a=4.707$ and $b=0.1282$; 
(ii)~$\chi^{2}_{fit}=0.296$; and (iii)~probability $P(\chi^{2}>\chi^{2}_{fit})
=0.9999$. Both fits are excellent and the power-law fit is consistent with
the result of Ref.~\cite{far1}.

\begin{figure}[p!]
\begin{center}
\mbox{
      \subfigure[\vspace{-0.2in}$\overline{P}=0.001$]{\label{fig2a}
            \includegraphics[scale=0.4]{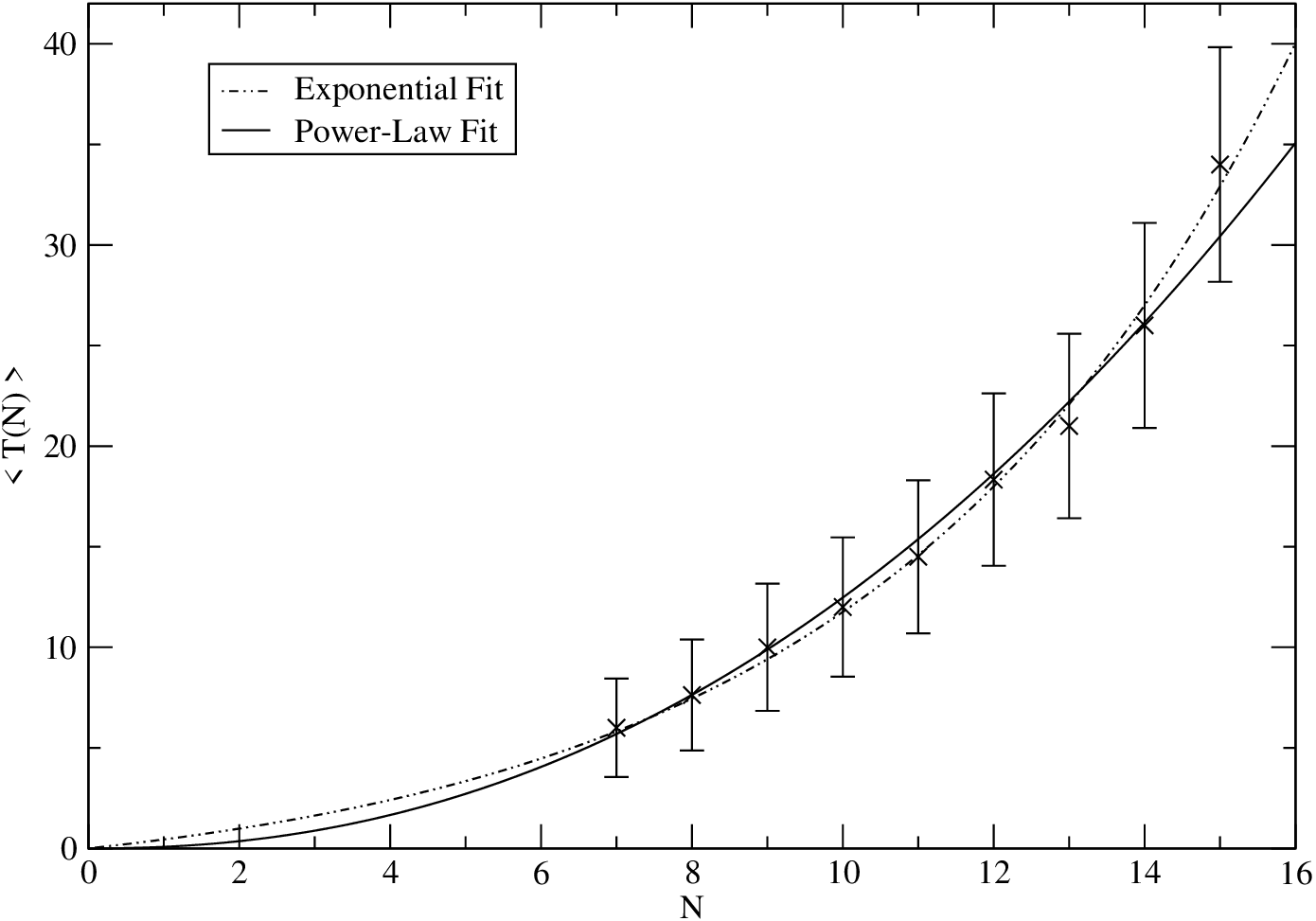}}
     }\\
      \vspace{0.5in}
\mbox{
      \subfigure[\vspace{-0.2in}$\overline{P}=0.003$]{\label{fig2b}
            \includegraphics[scale=0.4]{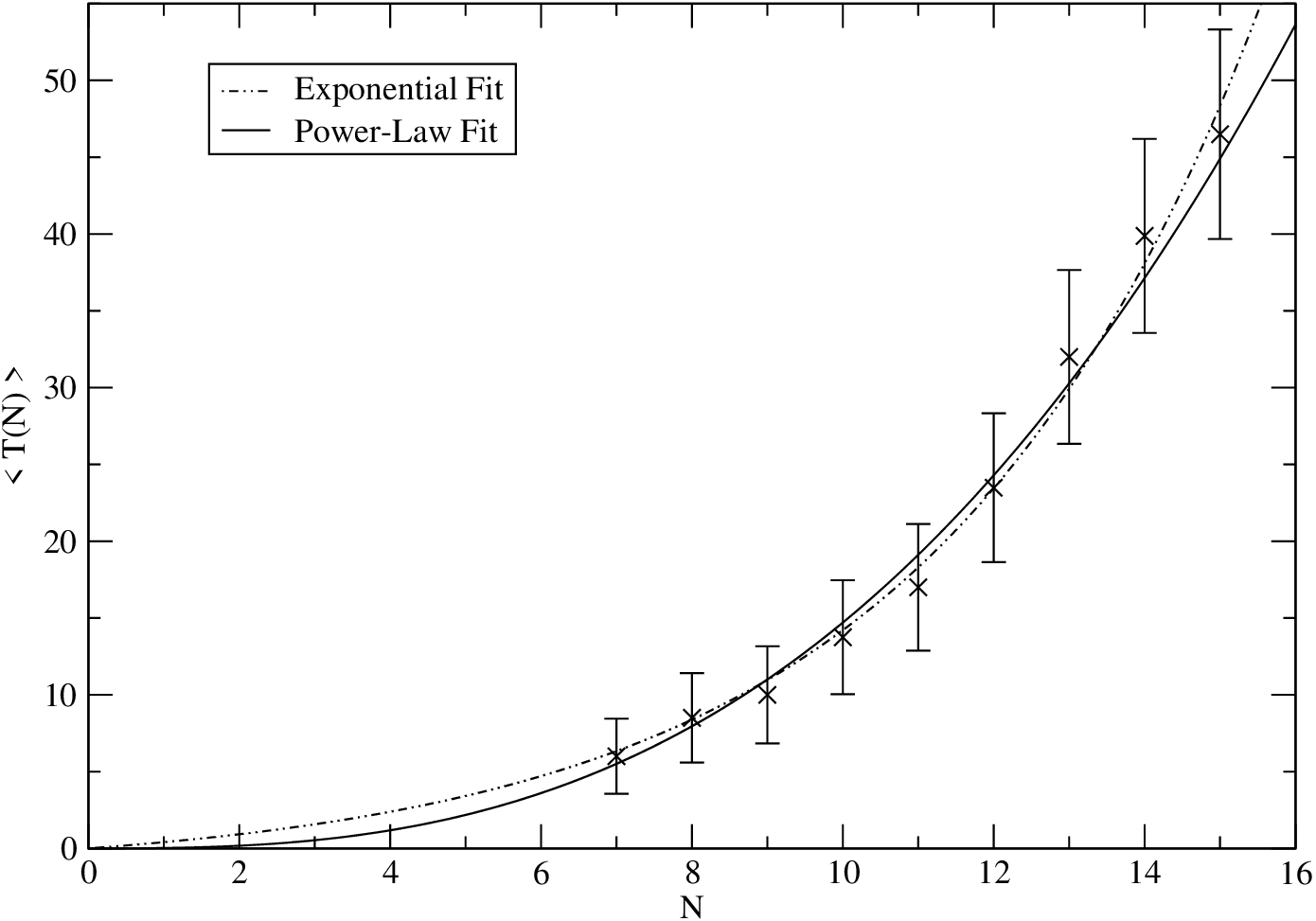}}
     }
\end{center}
\caption{\label{figure2}}
\end{figure}

\begin{figure}[p!]
\begin{center}
\mbox{
      \subfigure[\vspace{-0.2in}$\overline{P}=0.005$]{\label{fig3a}
            \includegraphics[scale=0.4]{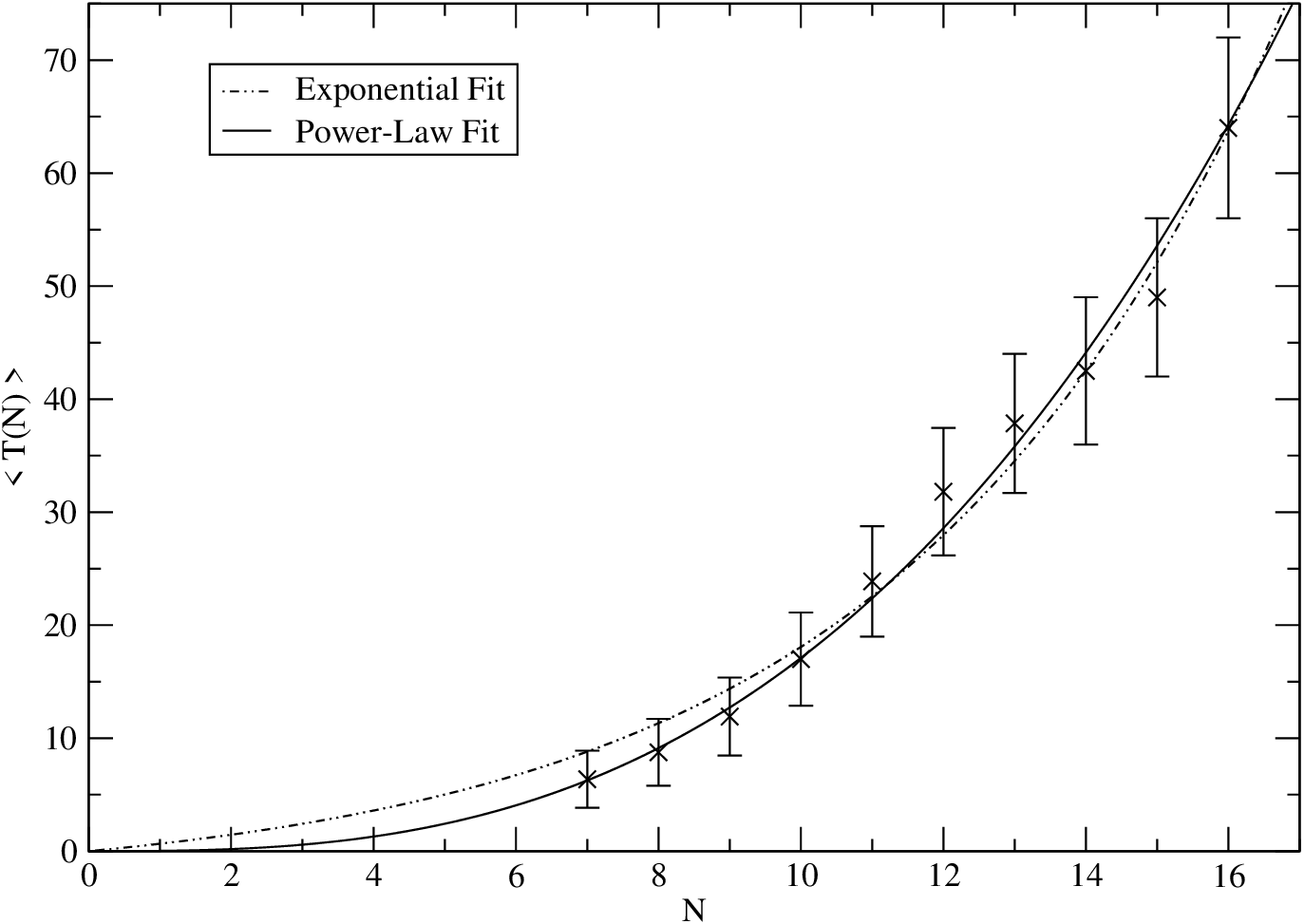}}
     }\\
      \vspace{0.5in}
\mbox{
      \subfigure[\vspace{-0.2in}$\overline{P}=0.007$]{\label{fig3b}
            \includegraphics[scale=0.4]{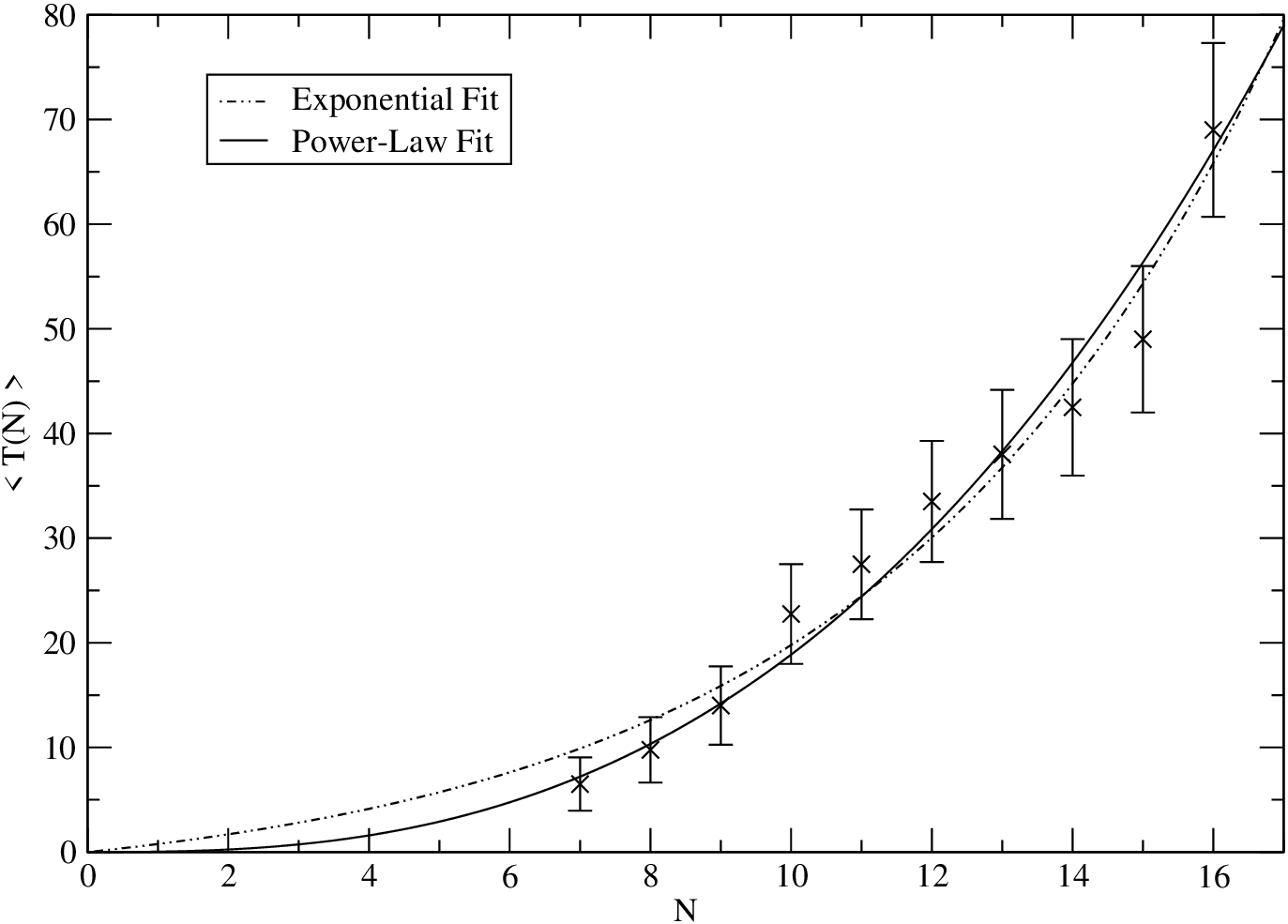}}
     }
\end{center}
\caption{\label{figure3}}
\end{figure}

\begin{figure}[p!]
\begin{center}
\mbox{
      \subfigure[\vspace{-0.2in}$\overline{P}=0.009$]{\label{fig4a}
            \includegraphics[scale=0.4]{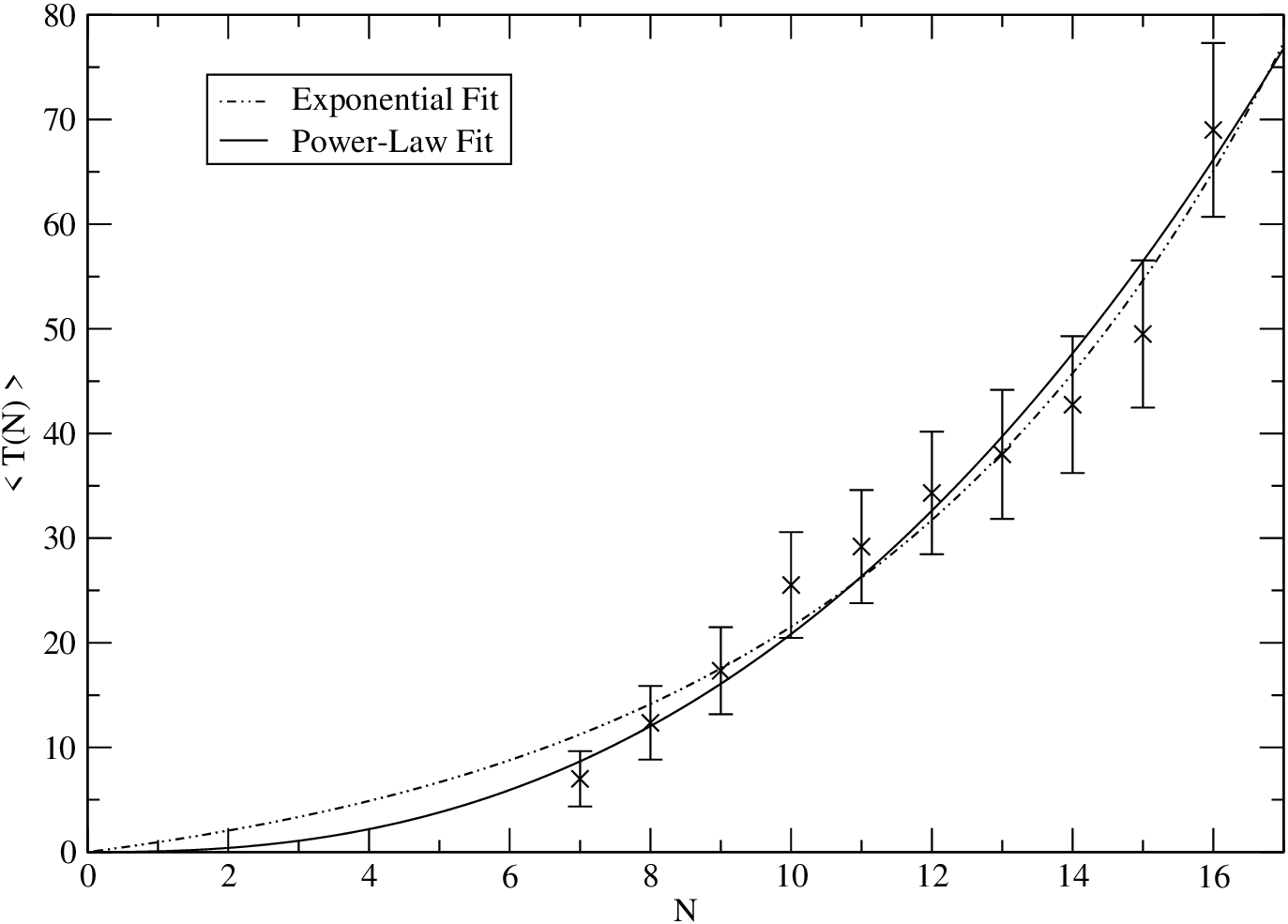}}
     }\\
      \vspace{0.5in}
\mbox{
      \subfigure[\vspace{-0.2in}$\overline{P}=0.013$]{\label{fig4b}
            \includegraphics[scale=0.4]{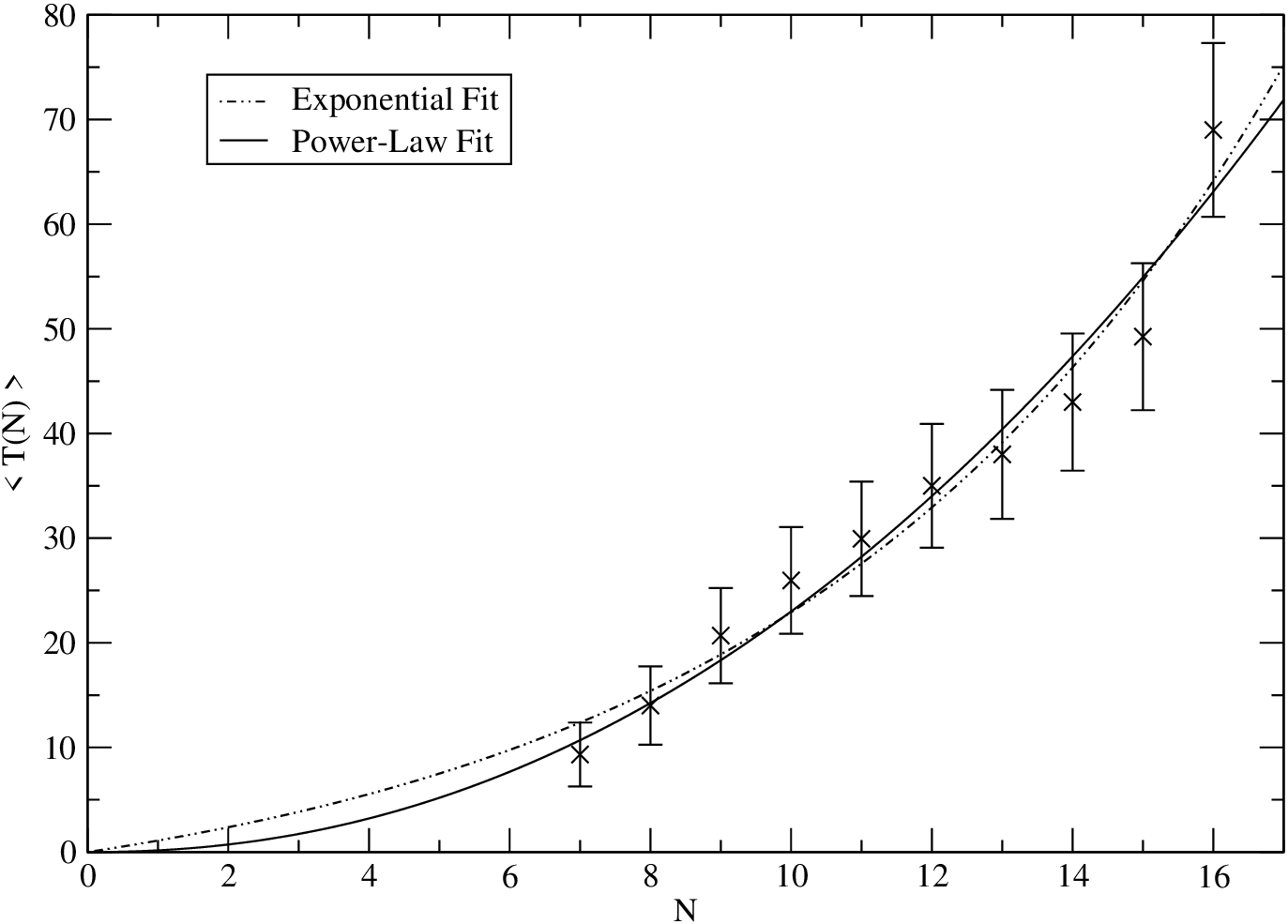}}
     }
\end{center}
\caption{\label{figure4}}
\end{figure}

The results for our noisy QuAdS simulations appear in $3$ figures and $2$
tables. Figures~\ref{figure2}, \ref{figure3} and \ref{figure4} plot the 
noise-averaged median runtime $\langle\overline{T}(N)\rangle$ versus $N$ for 
average noise power $\overline{P}=(0.001,0.003)$; $(0.005,0.007)$; and 
$(0.009,0.013)$, respectively. As in Figure~\ref{figure1}, each plot contains 
a power-law ($\langle \overline{T}(N)\rangle =aN^{b}$) and exponential 
($\langle\overline{T}(N)\rangle = a[\exp (bN)-1]$) fit to the simulation 
results. The parameters associated with the power-law and exponential fits 
appear, respectively, in Tables~\ref{table1} and \ref{table2}. For each value 
of $\overline{P}$ simulated, each Table contains: (i)~the best fit parameters 
$a$ and $b$; (ii)~the chi-squared for the fit $\chi^{2}_{fit}$; and (iii)~the 
probability $P(\chi^{2}>\chi^{2}_{fit})$. As noted above, the closer the 
latter probability is to $1$, the more consistent the data-set is with the 
fitting function. A discussion of these results is given in the following 
Section.

\section{Discussion}
\label{sec4}

Ref.~\cite{far3} examined the consequences of modifying the original linearly
interpolating QuAdS Hamiltonian $H(t)$ in eq.~(\ref{Hnon}) to 
\begin{equation}
\mathcal{H}(t) = H(t) + \delta\mathcal{H}(t) \hspace{0.1in} ,
\label{Hmod}
\end{equation}
where the new term $\delta\mathcal{H}(t)$ has the form
\begin{equation}
\delta\mathcal{H}(t) = e(t)\, H_{E} \hspace{0.1in} ,
\end{equation}
and the envelope function $e(t)$ is required to vanish at $t=0$ and $T$.
Ref.~\cite{far3} gave $3$ proposals for $H_{E}$, though it will not be 
necessary to review them here as we have a specific form in mind for 
$\delta\mathcal{H}(t)$ (see below). Eq.~(\ref{Hmod}) specifies a path in the 
space of $2^{N}\times 2^{N}$ Hermitian matrices that begins and ends at $H_{i}$
and $H_{P}$, respectively, and which by construction, differs from the linearly
interpolating path specified by $H(t)$. As noted in Section~\ref{sec1},
Ref.~\cite{far3} showed that by doing such a path variation, a failure of
QuAdS could be converted into a success.

Including a noise interaction in the QuAdS Hamiltonian also causes the
Hamiltonian path to deviate from the linearly interpolating path traced out
by  $H(t)$. For non-uniform $y$-polarized noise, the noise term in 
eq.~(\ref{Hint}) is
\begin{equation}
\delta\mathcal{H}(t) = -\sum_{j=1}^{N}\, N_{j}(t)\sigma_{y}^{j} 
\hspace{0.1in} ,
\end{equation}
where $\sigma_{y}^{j}$ is the Pauli matrix $\sigma_{y}$ for qubit $j$ and
$N_{j}(t)\,\hat{\mathbf{y}}$ is the noise field that interacts with this qubit.
Unlike the envelope function $e(t)$, the noise fields $\{ N_{j}(t):j=1,\ldots
, N\}$ need not vanish at $t=0$ or $T$. We now show, however, that for the
noise used in the simulations presented here and in Ref.~\cite{fg}, the
probability that a fluctuation is present at these times is small. To see this,
note that for a fluctuation $i$ to be  present at $t=0$ ($T$), the
fluctuation center $t_{i}$ must occur within a time $\tau$ of $t=0$ ($T$)
since the temporal width of the fluctuation is $2\tau$. For our noise model,
the time $t_{i}$ has a uniform probability distribution over the time interval
[$0,T$] so the probability that $t_{i}$ is within $\tau$ of $t=0$ ($T$) is
$\tau /T$. Since $\overline{n}$ is the average fluctuation rate, the average
number of fluctuations $\overline{N}_{\tau}$ that occur in a time interval
$\tau$ is $\overline{N}_{\tau} = \overline{n}\,\tau$. Since $\overline{n}=
\overline{P}/2\sigma^{2}\tau$ (see Section~\ref{sec2}), the total probability
that a fluctuation is present at $t=0$ ($T$) is
\begin{equation}
P_{tot} = \frac{\overline{P}\,\tau}{2\sigma^{2}T} \hspace{0.1in} .
\label{Ptot}
\end{equation}
For $\overline{P}=0.001$ ($0.013$), the midpoint for the range of $\langle
\overline{T}(N)\rangle$ values found in the simulation is approximately
$20$ ($39$). Using this value for $T$ in eq.~(\ref{Ptot}), and recalling that
our simulations used $\sigma =0.2$ and $\tau = 1$ gives $P_{tot}=6.25\times 
10^{-4}$ ($4.17\times 10^{-3}$) for $\overline{P}=0.001$ ($0.013$). Thus, with 
high probability, our noise interaction vanishes at $t=0$ ($T$), and our 
$\mathcal{H}(t)$ is equal to $H_{i}$ ($H_{P}$) at this time.

For non-uniform $y$-polarized noise, $\delta\mathcal{H}(t)$ has $N$ noise
fields $N_{j}(t)$, where $7\leq N\leq 16$ in our simulations. By comparison, 
the simulations in Ref.~\cite{fg} used uniform noise in which all qubits see 
the same noise field $N_{j}(t) =N(t)$, with $j=1,\ldots ,N$. Thus our noise
interaction with non-uniform $y$-polarized noise has an order of magnitude
more variation parameters than the uniform noise used in Ref.~\cite{fg}.
Furthermore, since the non-uniform noise simulations were done to larger values
of $\overline{P}$ than the simulations with uniform noise, each noise field 
in the former case has a larger average number of fluctuations 
$\overline{N}_{f}$ than in the latter case since $\overline{N}_{f}=\overline{P}
T/2\sigma^{2}\tau$. Thus, because the simulations in the present paper were
done using: (i)~non-uniform noise; (ii)~larger average noise power; and 
(iii)~larger number of qubits, they contain larger Hamiltonian path 
variations than was possible in Ref.~\cite{fg}. 

The simulation results presented in Tables~\ref{table1} and \ref{table2} show
that the scaling exponent $b$ begins to \textit{decrease\/} for 
$\overline{P}\geq 0.005$. (As will be discussed below, we anticipate that there
will be a maximum $\overline{P}$ value beyond which noise will begin to 
compromise QuAdS perfromance.) A second observation is that power-law scaling 
provides an excellent fit for all values of $\overline{P}$ simulated, while 
the exponential fit is not quite as good for $\overline{P}\geq 0.007$. To 
compare our results with those in Ref.~\cite{fg} a restricted power-law fit to 
the non-uniform noise results was done for $7\leq N\leq 12$ and $\overline{P}=
(0.001,0.003,0.005)$. This corresponds to the range of $N$ and $\overline{P}$ 
values simulated in Ref.~\cite{fg} for uniform $y$-polarized noise. The 
parameters for the restricted fit, together with the corresponding fit 
parameters for uniform $y$-polarized noise appear in Table~\ref{table3}. Note 
that a similar comparison is possible with exponential fits, though nothing 
new is learned and so we do not include that comparison here. From 
Table~\ref{table3} we see that the scaling exponent is comparable for the two
types of noise, with smaller $b$-values for non-uniform noise. Compared to the
$\overline{P}=0.000$ results, we see that noise slows down QuAdS, but 
Table~\ref{table3} gives a first indication that non-uniform noise may allow
conditions for the slowdown to be ameliorated. 

\begin{figure}[p!]
\centerline{\includegraphics[scale=0.65]{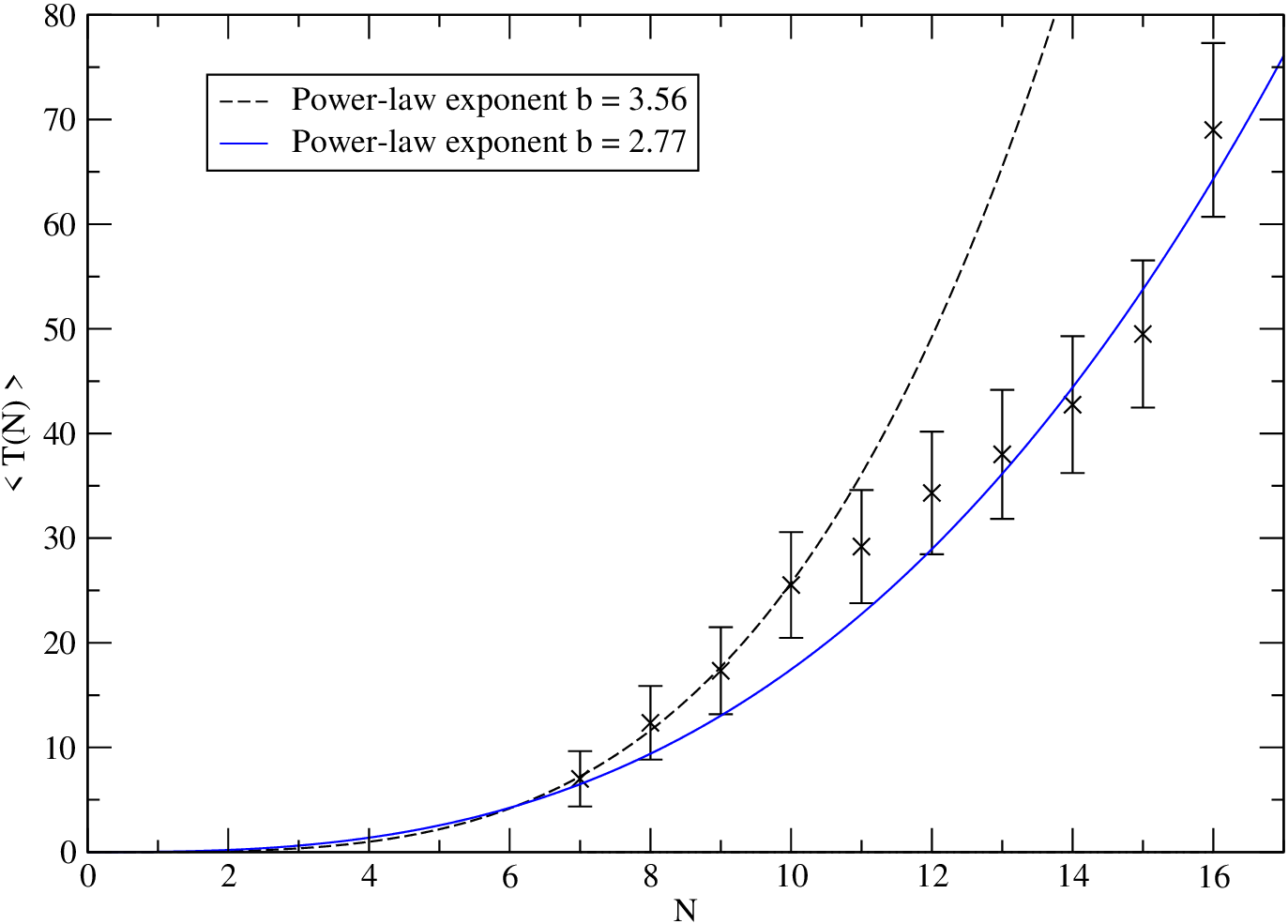}}
\vspace{3.5in}
\caption{\label{figure5}}
\end{figure}

A look at Figures~\ref{figure3} and \ref{figure4} shows that the initial rate
of growth of $\langle\overline{T}(N)\rangle$ appears to flatten out at
intermediate values of $N$. This flattening out is less pronounced for 
$\overline{P}=0.005$, occurring over the range $12\leq N\leq 14$; and is
broader for $\overline{P}=0.013$, occurring over the range $10-11\leq N\leq 
14$. To test this observation we did separate fits for each of the data-sets
in these Figures at small $N$ ($7\leq N\leq 10$) and (relatively) large $N$
($13\leq N\leq 16$). The parameters for the two fits appear in 
Table~\ref{table4}. As above, we only show results for a power-law fit.
Table~\ref{table4} indicates that the small $N$ fit grows at a faster rate
(viz.\ larger $b$) than the large $N$ fit. We see that the flattening out of
$\langle\overline{T}(N)\rangle$ at intermediate $N$ marks a crossover from 
rapid initial growth to a region of slower growth. In an effort to further
highlight this point, Figure~\ref{figure5} replots the data for $\overline{P}
=0.009$, including the fits for small and large $N$. One clearly sees the
data initially following the faster rising fit ($b=3.560$) and then crossing
over to the slower rising fit ($b=2.774$). Although noise is plainly causing 
QuAdS to slowdown relative to noiseless QuAdS (see Figure~\ref{figure1}),
Table~\ref{table4} indicates that non-uniform $y$-polarized noise can, for 
appropriate values of $\overline{P}$ and $N$, ameliorate the slowdown. One 
might wonder whether there is a connection between this ameliorating effect and
the suggestion made in Ref.~\cite{far3} that varying the Hamiltonian path away 
from the linearly interpolating path used in Refs.~\cite{far1} and \cite{far2} 
might improve QuAdS performance. From that perspective, one might wonder 
whether, for $12<N\leq 16$, non-uniform $y$-polarized noise with $0.005\leq
\overline{P}\leq 0.013$ is inducing sufficient variation of the linearly 
interpolating Hamiltonian path to yield alternative paths with slightly larger 
minimum energy gaps, causing a (slightly) improved adiabaticity, and so 
reducing (slightly) the noise-induced slowdown of QuAdS seen at smaller $N$ 
values. A proper examination of this 
ansatz requires that the minimum gaps be determined for the USA instances 
and noise realizations that we simulated in this paper. We plan to carry out
this analysis in future work. Note that for sufficiently large average noise 
power, noise-induced decoherence should ultimately rob QuAdS of its quantum 
performance-enhancements since it will cause the dynamics to crossover from 
quantum to classical. A quantitative determination of how this loss of quantum 
performance occurs presents a significant (though fascinating) challenge for 
future simulations. 

\section*{Acknowledgments}
We thank the NSF Cyberinfrastructure Partnership for access to the TeraGrid
cluster through a Large Resource Allocation (grant MCA05T020T) and T. Howell
III for continued support.

\pagebreak
\begin{center}
\textbf{Figure Captions}
\end{center}
\begin{description}
\item[Figure 1:] Noiseless QuAdS simulation results for the median runtime
$\overline{T}(N)$ (dimensionless units) versus number of bits $N$. The solid 
line is the best power-law fit to the data and the dash-dot line is the best 
exponential fit. The error bars give $95\%$ confidence limits on the median.
\item[Figure 2:] Simulation results for the noise-averaged median runtime
$\langle\overline{T}(N)\rangle$ (dimensionless units) versus number of bits
$N$. The noise is polarized along $\hat{\mathbf{y}}$ and has average noise
power $0.001$ and $0.003$ (dimensionless units). Each datapoint is the median
of $750$ runtimes ($75$ USA instances and $10$ noise environments per USA 
instance). The solid-line is the best power-law fit to the data and the
dash-dot line is the best exponential fit. The error bars give $95\%$ 
confidence limits for each median.
\item[Figure 3:] Simulation results for the noise-averaged median runtime
$\langle\overline{T}(N)\rangle$ (dimensionless units) versus number of bits
$N$. The noise is polarized along $\hat{\mathbf{y}}$ and has average noise
power $0.005$ and $0.007$ (dimensionless units). Each datapoint is the median
of $750$ runtimes ($75$ USA instances and $10$ noise environments per USA 
instance). The solid-line is the best power-law fit to the data and the
dash-dot line is the best exponential fit. The error bars give $95\%$ 
confidence limits for each median.
\item[Figure 4:] Simluation results for the noise-averaged median runtime
$\langle\overline{T}(N)\rangle$ (dimensionless units) versus number of bits
$N$. The noise is polarized along $\hat{\mathbf{y}}$ and has average noise
power $0.009$ and $0.013$ (dimensionless units). Each datapoint is the median
of $750$ runtimes ($75$ USA instances and $10$ noise environments per USA 
instance). The solid-line is the best power-law fit to the data and the
dash-dot line is the best exponential fit. The error bars give $95\%$ 
confidence limits for each median.
\item[Figure 5:] Re-plot of the simulation results for $\overline{P}=0.009$.
The dashed (solid) curve is a restricted power-law fit $\langle\overline{T}(N)
\rangle = aN^{b}$ through the data $7\leq N\leq 10$ ($13\leq N\leq 16$).
Similar plots are possible for the other $\overline{P}$--values in 
Table~\ref{table4}, though we do not include them to avoid repetition.
\end{description}
\begin{table}[p!]
\vspace{0.25in}
\caption{\label{table1}Best fit parameters for power-law scaling $\langle
\overline{T}(N)\rangle =aN^{b}$ for all values of average noise power
$\overline{P}$ simulated. For comparison, best-fit parameters for noiseless
QuAdS are also included.}
\begin{center}
\begin{tabular}{|c|c|c|c|c|}\hline\hline    
\rule{0mm}{4mm}$\overline{P}$ & $a$ & $b$ & $\chi^{2}_{fit}$ & $P(\chi^{2}>
\chi^{2}_{fit})$\\\hline
\rule{0mm}{4mm}$0.000$ & $0.1016$ & $2.079$ & $0.321$ & $0.9999$ \\
\rule{0mm}{5mm}$0.001$ & $0.07863$ & $2.200$ & $0.539$ & $0.9993$ \\
$0.003$ & $0.02595$ & $2.753$ & $0.869$ & $0.9967$ \\
$0.005$ & $0.02611$ & $2.816$ & $1.097$ & $0.9976$ \\
$0.007$ & $0.03783$ & $2.698$ & $2.919$ & $0.9393$ \\ 
$0.009$ & $0.07251$ & $2.458$ & $3.449$ & $0.9032$ \\ 
$0.013$ & $0.1636$ & $2.148$ & $2.689$ & $0.9523$ \\\hline
\end{tabular}
\end{center}
\end{table}
\begin{table}[p!]
\caption{\label{table2}Best fit parameters for exponential scaling $\langle
\overline{T}(N)\rangle =a[\exp (bN) - 1]$ for all values of average noise power
$\overline{P}$ simulated. For comparison, best-fit parameters for noiseless
QuAdS are also included.}
\begin{center}
\begin{tabular}{|c|c|c|c|c|}\hline\hline    
\rule{0mm}{4mm}$\overline{P}$ & $a$ & $b$ & $\chi^{2}_{fit}$ & $P(\chi^{2}>
\chi^{2}_{fit})$\\\hline
\rule{0mm}{4mm}$0.000$ & $4.707$ & $0.1282$ & $0.296$ & $0.9999$ \\
\rule{0mm}{5mm}$0.001$ & $2.208$ & $0.1845$ & $0.185$ & $0.9999$ \\
$0.003$ & $1.580$ & $0.2302$ & $0.517$ & $0.9994$ \\
$0.005$ & $3.123$ & $0.1915$ & $3.307$ & $0.9136$ \\
$0.007$ & $3.924$ & $0.1799$ & $4.833$ & $0.7753$ \\ 
$0.009$ & $5.445$ & $0.1601$ & $4.927$ & $0.7653$ \\ 
$0.013$ & $7.112$ & $0.1440$ & $3.172$ & $0.9231$ \\ \hline 
\end{tabular}
\end{center}
\end{table}
\begin{table}[p!]
\caption{\label{table3}Comparison of power-law fit $\langle\overline{T}(N)
\rangle =aN^{b}$ for non-uniform and uniform (Ref.~\cite{fg}) $y$-polarized
noise for $7\leq N\leq 12$ and average noise power $\overline{P}=0.001$,
$0.003$, and $0.005$.}
\begin{center}
\begin{tabular}{|c|cc|cc|}\hline\hline
\rule[-2mm]{0mm}{6mm}$\overline{P}$ & \multicolumn{2}{c|}{non-uniform}
   & \multicolumn{2}{c|}{uniform} \\ \hline
\rule{0mm}{4mm}$-$ & $a$ & $b$ & $a$ & $b$ \\ \hline
\rule{0cm}{4mm}$0.001$ &$1.107\times 10^{-1}$ & $2.044$ &
   $9.677\times 10^{-2}$ & $2.108$ \\
$0.003$ & $5.095\times 10^{-2}$ & $2.439$ & $2.982\times 10^{-2}$ &
   $2.679$\\
$0.005$ & $1.677\times 10^{-2}$ & $3.020$ & $6.607\times 10^{-3}$ &
   $3.429$\\ \hline
\end{tabular}
\end{center}
\end{table}
\begin{table}[p!]
\caption{\label{table4}Restricted power-law fits $\langle\overline{T}(N)\rangle 
= aN^{b}$ for the $\overline{P}=0.005,0.007,0.009,0.013$ data-sets for:
(i)~$7\leq N\leq 10$; and (ii)~$13\leq N\leq 16$.}
\begin{center}
\begin{tabular}{|c|cc|cc|}\hline\hline
\rule[-2mm]{0mm}{6mm}$\overline{P}$ & \multicolumn{2}{c|}{$7\leq N\leq 10$}
 & \multicolumn{2}{c|}{$13\leq N\leq 16$} \\ \hline
\rule{0mm}{4mm}$-$ & $a$ & $b$ & $a$ & $b$ \\ \hline
\rule{0cm}{4mm}$0.005$ & $3.064\times 10^{-2}$ & $2.730$ &
   $6.554\times 10^{-2}$ & $2.465$ \\
$0.007$ & $7.637\times 10^{-3}$ & $3.450$ & $2.984\times 10^{-2}$ &
   $2.767$\\
$0.009$ & $7.086\times 10^{-3}$ & $3.560$ & $2.934\times 10^{-2}$ &
   $2.774$\\
$0.013$ & $3.201\times 10^{-2}$ & $2.924$ & $3.060\times 10^{-2}$ &
   $2.759$\\ \hline
\end{tabular}
\end{center}
\end{table}
\pagebreak
\end{document}